\documentclass[twocolumn,times]{aastex631}

\usepackage{amsmath}
\usepackage{bm}

\def\ihep{Key Laboratory for Particle Astrophysics, Institute of High Energy Physics,
Chinese Academy of Sciences, 19B Yuquan Road, Beijing 100049, China;
\href{mailto:wangjm@mail.ihep.ac.cn}{wangjm@mail.ihep.ac.cn},
\href{mailto:liyanrong@ihep.ac.cn}{liyanrong@ihep.ac.cn}}

\def\UCASastro{School of Astronomy and Space Sciences, University of Chinese Academy of Sciences, 
19A Yuquan Road, Beijing 100049, China}

\def\UCASphy{School of Physical Sciences, University of Chinese Academy of Sciences, 
19A Yuquan Road, Beijing 100049, China}

\def\NAOC{National Astronomical Observatory of China, 20A Datun Road, Beijing 100020, China}

\begin{document}
\title{\large A New Interpretation for the Hot Corona in Active Galactic Nuclei}

\author[0009-0006-7629-1459]{Yu Zhao}
\affil{\ihep}
\affil{\UCASphy}

\author[0000-0001-5841-9179]{Yan-Rong Li}
\affil{\ihep}

\author[0000-0001-9449-9268]{Jian-Min Wang}
\affil{\ihep}
\affil{\UCASastro}
\affil{\NAOC}

\def\LSUN{\rm L_{\odot}}
\def\MSUN{\rm M_{\odot}}
\def\RSUN{\rm R_{\odot}}
\def\MSUNYR{\rm M_{\odot}\,yr^{-1}}
\def\MSUNS{\rm M_{\odot}\,s^{-1}}
\def\MDOT{\dot{M}}

\begin{abstract}
This work attempts to provide a new interpretation for the hot corona in active galactic nuclei (AGNs).  
A thin parabolic magnetic reconnection layer, anchored at the innermost disk and extending along the boundary of the magnetic tower for a few tens of gravitational radii, serves as a hard X-rays source above the disk. Within this reconnection layer, the tearing instability leads to formation of a chain of plasmoids, which contain relativistic electrons that generate X-ray radiation through inverse Compton (IC) scattering of soft photons emitted by the accretion disk. Based on previous theoretical works and numerical simulations, we develop a heuristic framework to parameterize the geometry and magnetization of the reconnection layer, as well as to compute both the power of IC scattering radiation and the height of the reconnection layer. Our model allows for a quantitative investigation of the relation between height of the corona and the X-ray radiation luminosity, which can be directly compared against the observed relation from X-ray reverberation mapping of individual AGNs. The theoretical results are in good agreement with the observations of IRAS 13224-3809, indicating the validation of our model.
\end{abstract}


\section{INTRODUCTION}
Active galactic nuclei (AGNs) have been intensively studied across most (if not all) of the electromagnetic spectrum, from the radio, optical, ultraviolet (UV), and X-ray to gamma-ray wave bands. It is widely accepted that the radio and gamma-ray emissions predominantly arise from AGN jets, while the optical and UV radiation is associated with the accretion disk. In the hard X-ray regime, observations show that a power-law shape with a high-energy cutoff provides a good fit for both radio-quiet and radio-loud AGNs (\citealt{Ricci2011,Molina2013,Soldi2014}). This plausibly implies a unified mechanism for hard X-ray emissions in AGNs. A common suggestion posits that the hard X-ray radiation originates from the inverse-Compton (IC) scattering of disk seed photons by relativistic electrons
within a hot corona. This corona is characterized by an electron temperature of $kT_{\text{e}}\sim\text{100~keV}$ and a Thomson-scattering optical depth of $\tau_{\text{T}}\sim 1$, located within a few tens of gravitational radii from the central black hole (BH; \citealt{Fabian2015,Wilkins2015}).
Although such a hot-corona picture has long been proposed (\citealt{Haardt1991}), hitherto the hot-corona models in the literature have still been largely phenomenological, and the underlying nature of the hot corona in terms of its geometry, kinematics, and formation remains elusive.

Regarding the geometry and dynamics of the hot corona, observations have shown that the hard X-rays of AGNs exhibit minute timescale variations \citep{Vaughan2011,Alston2019}, indicative of the compact size of the corona. 
Furthermore, utilizing spectral analysis, \cite{Pal2023} and \cite{Serafinelli2024} found that variations in temperature and optical depth of the corona do not show a significant correlation with X-ray variations in both short timescales of hours and long timescales from days to years. They thereby proposed that X-ray variations are more likely driven by changes in the corona's size and/or geometry.
 The lamppost model, commonly utilized in reverberation mapping of AGNs, assumes that the hot corona is a static, point-like source that emits isotropic power-law X-rays, located at a fixed height of several gravitational radii above the BH along the rotation axis \citep{Alston2020}. Thus, the hot corona in reverberation mapping is primarily characterized by two parameters: its height and luminosity. 
However, some studies have indicated that these assumptions about the hot corona in the lamp-post model may not be accurate.
For instance, in order to explain the observation that radio-quiet AGNs exhibit stronger X-ray reflection compared to radio-loud AGNs (\citealt{Zdziarski1995,Wozniak1998,Eracleous2000}), it has been proposed that the hot corona is outflowing at mild relativistic speeds (\citealt{Beloborodov1999,Malzac2001}). Additionally, \cite{Liu2014} found that type 1 AGNs in the radio-quiet category are intrinsically brighter in the 2-10 keV band compared to type 2 AGNs by a factor of 2.8, which can be attributed to the beaming effect of an outflowing corona moving at velocities between $0.3\sim0.5$ times the speed of light. These observations indicate that the hot corona is neither static nor isotropic. 
Reverberation mapping in X-rays pointed to a more dynamic and complex geometry for the hot corona: \citet{Alston2020} found that the hot corona height in IRAS 13224-3809 is variable and positively correlated to its radiation power in 2-10 keV band. \citet{Wilkins2023} explored variations of the corona height on short timescales using spectral timing analyses. Collectively, these studies illustrate that the hot corona exhibits more intricate geometry and dynamics than traditionally assumed in the lamp-post model.

An essential ingredient for understanding formation of the hot corona is the origin of relativistic electrons. Magnetic reconnection has been suggested as a potential mechanism for accelerating electrons, which transfers magnetic energy to the kinetic energy of particles, ultimately leading to IC scattering radiation within a reconnection layer. In observational aspects, magnetic reconnection with MHD turbulence offers an explanation for the luminosity and short-timescale variability of X-rays in AGNs  \citep{Matteo1998}. Additionally, the strong correlation between radio and X-ray luminosity observed in radio-quiet AGNs has also been ascribed to magnetic reconnection in the hot corona (e.g. \citealt{Panessa2019}).
Moreover, when the reconnection layer becomes sufficiently thin, a chain of moving plasmoids will be formed by tearing instability (\citealt{Loureiro2007,Uzdensky2010,Sironi2016,Sironi2020,Sridhar2021,Sridhar2023}), which has been used to explain the observed power-law X-ray spectrum and X-ray flares (\citealt{Beloborodov2017}). Furthermore, general relativistic magnetohydrodynamics (GRMHD) simulations indicate that such a chain of plasmoids can develop along a jetlike sheath, extending at least to a few tens of gravitational radii (\citealt{Parfrey2015,Ripperda2020,Davelaar2023}), thereby physically presenting a vertically extended hard-X-ray source. 

Inspired by these factors, we attempt to link the physical characteristics of the
hot corona with this vertically extended reconnection layer.
To this end, it is necessary to clarify the possibility of the presence of a large-scale magnetic field in the innermost region of AGNs. In relation to theoretical and numerical simulations, previous works have argued that the large-scale bipolar field will be dragged inward as matter accretes and then accumulates in the innermost region of the disk, rather than being accreted into the BH or locally dissipated by outward magnetic diffusion (\citealt{Igumenshchev2003,Spruit2005,Bisnovatyi-Kogan2007,Lovelace2009,McKinney2012,Cao2013,Igumenshchev2008}). This advection of the large-scale magnetic field is efficient in a radiatively inefficient accretion flow \citep{Narayan1995,Cao2011} or radiatively efficient accretion flow with magnetically driven outflows \citep{Cao2013}.
In observations, the presence of jets often signifies a strong magnetic field, as the magnetic flux threading the BH is believed to be a necessary condition for jet formation \citep{Sikora2013}. 
Optical spectropolarimetry observations have suggested a strong magnetic field near the BH horizon in both radio-loud and radio-quiet AGNs \citep{Piotrovich2021}.
If the detected ionized outflows (\citealt{Laha2021}) are driven by magnetohydrodynamic (MHD) processes \citep{Blandford1982}, this will further imply the existence of a large-scale magnetic-field-threading accretion disk.  
Therefore, we hypothesize that a large-scale magnetic field can accumulate in the innermost disk, enabling magnetic reconnection to occur and form a reconnection layer. 

This work makes an attempt at exploring the physical nature of the hot corona above the disk. 
We propose a physical and quantitative model for the hot corona in AGNs by investigating the geometry and radiation properties of a magnetic reconnection layer.  
 The paper is organized as follows. 
 In Section \ref{field_rec}, we introduce a magnetic field configuration for the generation of a plasmoid chain within a reconnection layer. 
 In Sections \ref{geometry}-\ref{radiation}, we analyze the geometry and radiation of the reconnection layer. 
 In Section \ref{application}, we apply our model to IRAS 13224-3809 and compare with observations. 
 Finally, we summarize our work and provide discussions of several key issues in Section \ref{discussion}.

\begin{figure*}[ht!]
\centering
\includegraphics[width=0.8\textwidth]{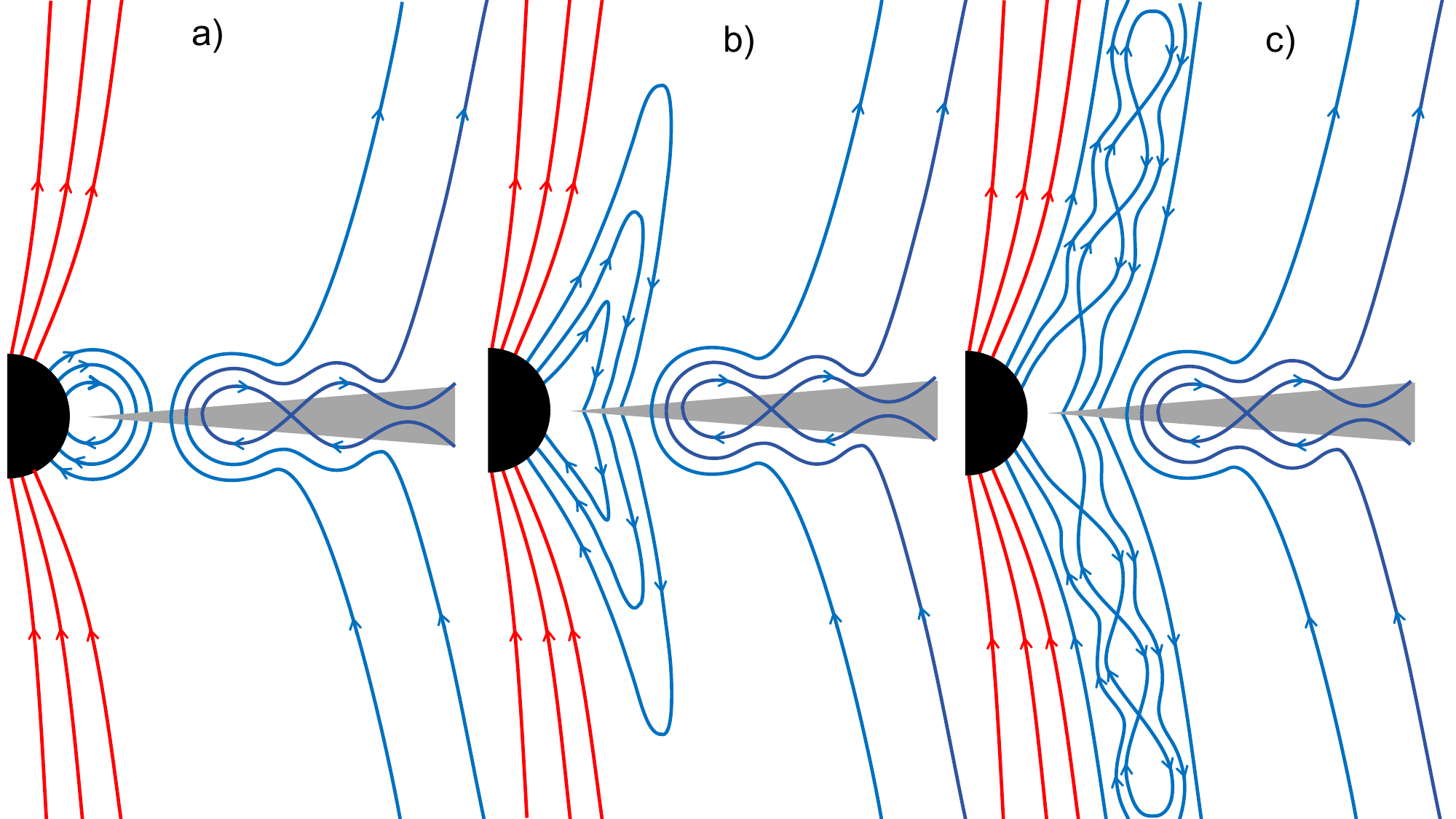}
\caption{A cartoon diagram of the magnetic field configuration in a BH-disk system. The red lines represent the open magnetic field lines threading the BH, while the blue lines represent the field lines of the magnetic tower and accretion disk. The black area represents the BH, and the gray area represents the accretion disk. (a) A magnetic loop disconnects from the large-scale field and is advected to attach to the BH, with one footpoint connecting to the horizon. 
(b) The magnetic loop is twisted and powered by the differential rotation between its footpoints, then opens up to stretch along the jetlike sheath, forming an elongated current sheet. 
 (c) This thin current sheet is tearing-unstable, leading to the formation of a chain of plasmoids. 
 The twists of the field lines in the tower resulting from the differential rotation are not shown.
 }
\label{Fig_field_config}
\end{figure*}

\section{THEORETICAL MODEL AND APPLICATION}
First, we emphasize some basic quantities of a BH-disk system in cylindrical polar coordinates $(R,\phi, Z)$, where $R$ and $Z$ represent radius and height, respectively, and $\phi$ denotes the azimuth. 
For a BH with a mass of $M$ and spin parameter of $a$, the radius of the BH horizon is given by
$R_{\rm H}=R_{\rm g}\left(1+\sqrt{1-a^{2}}\right)$, where $R_{\rm g}=GM/c^{2}$ is the gravitational radius of the BH, $G$ is the gravitational constant and $c$ is the speed of light. The innermost boundary of an accretion disk is commonly considered to be at the radius of the innermost stable circular orbit, which reads $R_{\rm in}= R_{\rm g}\left[3+Z_{2}\mp\sqrt{(3-Z_{1})(3+Z_{1}+2Z_{2})}\right]$ (see, e.g., \citealt{Wald1984}),
where the negative and positive signs correspond to $a > 0$ and $a < 0$, respectively, and
$Z_{1}=1+\left(1-a^{2}\right)^{1/3}\left[(1+a)^{1/3}+(1-a)^{1/3}\right]$,
$Z_{2}=\sqrt{3a^{2}+Z_{1}^{2}}$. For the convenience of the calculations, we introduce the following dimensionless quantities $m=M/M_{\odot}, \dot{m}=\dot{M}/\dot{M}_{\rm Edd}, r=R/R_{\rm g}$, where the critical accretion rate $\dot{M}_{\rm Edd}=L_{\rm Edd}/c^{2}=1.4\times 10^{17}m\cdot\rm g~s^{-1}$, $L_{\rm Edd}= 4\pi cGMm_{\rm H}/\sigma_{\rm T}$ is the Eddington luminosity, $m_{\rm H}$ is the mass of hydrogen atoms, and $\sigma_{\rm T}$ is the cross section of Thomson scattering. 

\subsection{Field Configuration for the Plasmoid Chain} \label{field_rec}
If the large-scale magnetic field can be effectively advected into the BH, as discussed above, a phenomenological and simplified magnetic field configuration inferred from theories and numerical simulations can be constructed (see Figure \ref{Fig_field_config}). 
The accretion flow can be treated as a series of azimuthally axisymmetric current rings, generating large-scale poloidal loops that thread through the disk, which account for the production of magnetic outflows \citep{Blandford1982,Spruit1996}. 
These magnetic loops are advected with the accretion flow from the outer region of the disk to the inner region.
In Figure. \ref{Fig_field_config}(a), as the innermost magnetic loops reach the innermost region and approach the BH, they may disconnect from the large-scale field through reconnection and then attach to the horizon (for details, refer to chapter 8.1 of \citealt{Punsly2001}). 
Open poloidal field lines threading the BH and connecting to infinity are commonly suggested (\citealt{Blandford1977, Bi2007, Blandford2022}), but the presence of closed field lines with one footpoint connecting to the BH and the other to the disk has also been proposed (\citealt{de2005,Blandford2022}).  
Specifically, if there is differential rotation between the footpoints of these closed field loops, they are unable to co-rotate rigidly with the disk and BH. They will be twisted and powered to generate a toroidal field component that produces a vertical pressure gradient force, causing the field lines to rise away from the disk. This eventually leads to the formation of a helical magnetic structure that can extend along a vertical sheath. 
This specific field configuration has been proposed to elucidate the collimation of jets  (\citealt{Lynden-Bell1996,Lynden-Bell2003,Lynden-Bell2006}), known as the magnetic tower, which has been verified through MHD simulations \citep{Kato2004a,Kato2004b} and observed in the M87 jet \citep{Pasetto2021}. As these extended flux loops open up, this can trigger the tearing instability and lead to the formation of a chain of plasmoids in a thin reconnection layer \citep{Loureiro2007}, as depicted in Figure \ref{Fig_field_config}c. Simulations shows that such a reconnection layer is primarily caused by the differential rotation between the BH and the disk, which is anchored at the innermost region of the disk  (\citealt{Parfrey2015,Ripperda2020}). \citet{Beloborodov2017} explored the radiation properties of the plasmoid chain through theoretical analysis and Monte-Carlo simulations and found that those plasmoids provide a good explanation for the hard X-ray spectrum of AGNs. Additionally, the chain of plasmoids along the boundary of the magnetic tower has been observed in GRMHD simulations and employed to explain the observed X-ray flares (\citealt{Ripperda2020}). 

Based on this field configuration, we try to investigate whether the chain of plasmoids in the reconnection layer along the tower's boundary can be considered as the fundamental physical nature of the hot corona in AGN reverberation mapping. In reverberation mapping, the key quantities of the hot corona are its height and luminosity, so our aim is to explain these two quantities.

 \begin{figure*}[th!]
\centering 
\scalebox{0.5}[0.5]{\rotatebox{0}{\includegraphics{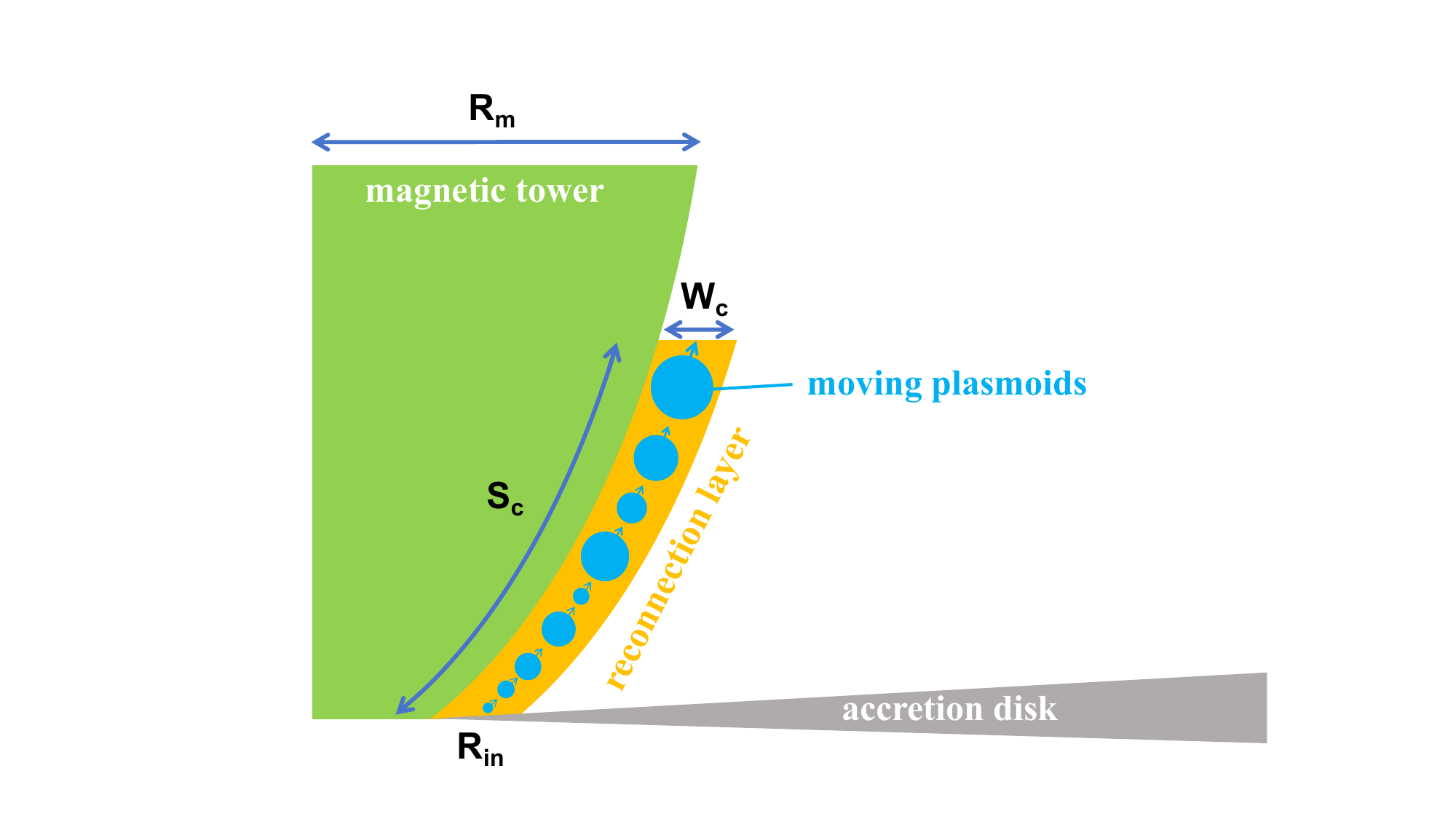}}}
\caption{A cartoon diagram of the reconnection layer configuration along the boundary of the magnetic tower. The gray area represents the accretion disk, the green area represents the force-free magnetic tower, the orange area represents the reconnection layer located at $R_{\rm in}$ with the characteristic curve length $S_{\rm c}$ and thickness $W_{\rm c}$, and the blue dots represent the relativistic plasmoids moving in the reconnection layer. The largest plasmoids have a typical size of $W_{\rm max}\sim W_{\rm c}$, while the smallest plasmoids have a size comparable to the Larmor radius, with $W_{\rm min}\sim R_{\rm L}$.}
\label{Fig_layer_config}
\end{figure*}

\begin{deluxetable}{ccc}
\tablecolumns{3}
\tabletypesize{\footnotesize}
\tabcaption{Parameter Sets of ($\alpha, \beta$) for the Shapes of the Magnetic Tower in Equation~(\ref{shape}). \label{tab_parameter}}
\tablehead{
\colhead{~~~~~~~~~~~~~~Set~~~~~~~~~~~~~~}&
\colhead{~~~~~~~~~~~~~~$\alpha$~~~~~~~~~~~~~~}&
\colhead{~~~~~~~~~~~~~~$\beta$~~~~~~~~~~~~~~}}
\startdata
    M1  & 0.1 & 2.0 \\
    M2  & 0.1 & 3.0 \\
    M3  & 0.1 & 4.0 \\
    M4  & 0.2 & 2.0 \\
    M5  & 0.2 & 3.0 \\
    M6  & 0.2 & 4.0 \\
    M7  & 0.3 & 2.0 \\
    M8  & 0.3 & 3.0 \\
    M9  & 0.3 & 4.0 
\enddata
\end{deluxetable}

\begin{figure*}[th!]
\centering 
\scalebox{0.6}[0.6]{\rotatebox{0}{\includegraphics{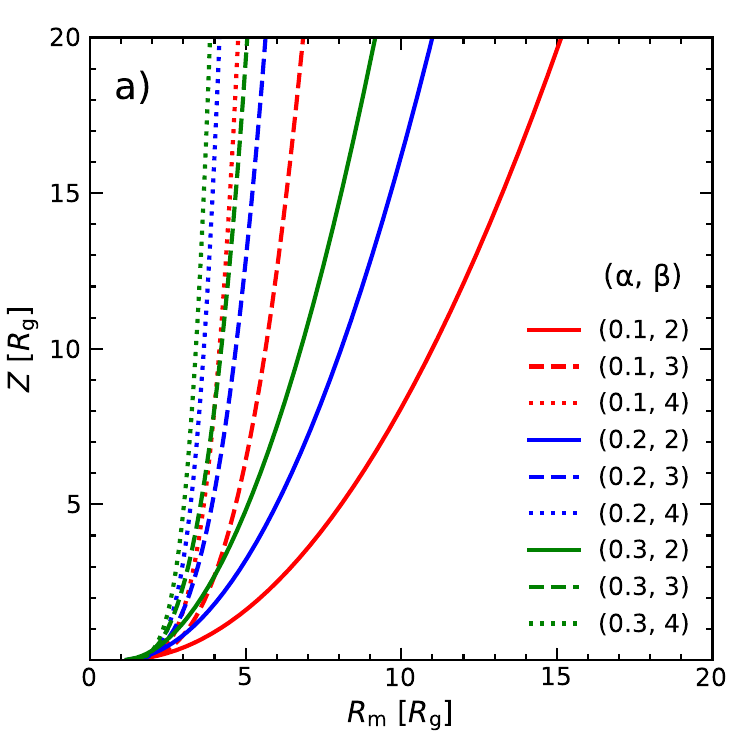}}}
\scalebox{0.6}[0.6]{\rotatebox{0}{\includegraphics{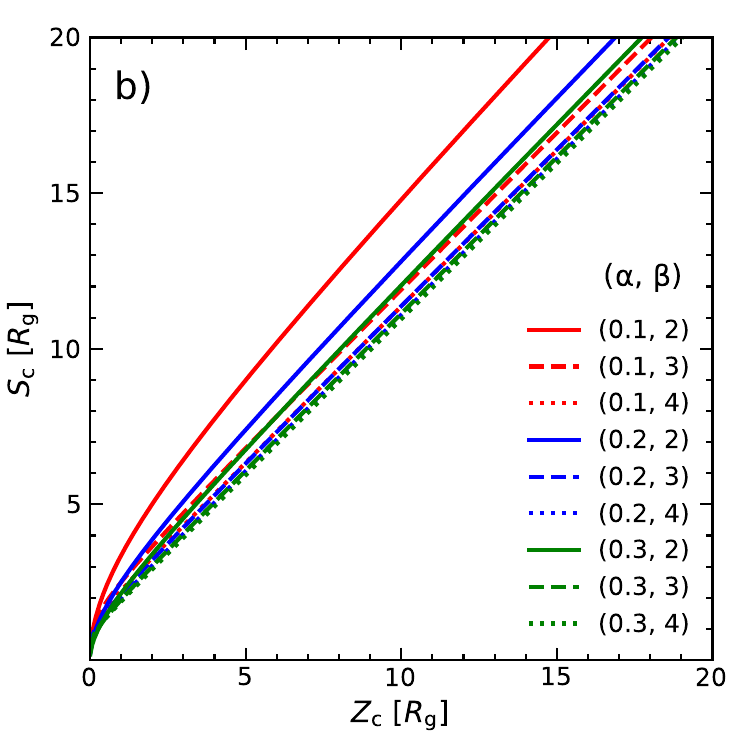}}}
\caption{(a) The shapes of the magnetic tower's boundary (i.e., the reconnection layers) for the different pairs of $\alpha$ and $\beta$ listed in Table~\ref{tab_parameter}. (b) The corresponding relations between the reconnection layer's curve length $S_{\rm c}$ and height $Z_{\rm c}$.}
\label{Fig_Z_R}
\end{figure*}

\subsection{The Geometry of the Reconnection Layer} \label{geometry}

As the chain of plasmoids forms in the reconnection layer along the boundary of magnetic tower, we analyze the geometry of the reconnection layer from the model of the magnetic tower proposed by \cite{Lynden-Bell2003}.
According to this model, given the magnetic field strength at the tower base, differential rotation of field lines, and external pressure, the shape of the tower's boundary can be self-consistently determined. This shape is described by the function $R_{\rm m}(Z)$, where $R_{\rm m}$ is the radius of the tower's cross-section area at height $Z$ as shown in Figure \ref{Fig_layer_config}.

For a tall tower, the tower shape follows a power-law dependence on pressure, as $R_{\rm m}\propto p_{\rm ext}^{-1/4}$ (see \citealt{Lynden-Bell2006} for details), where $p_{\rm ext}$ represents the external pressure. This relation can be understood by pressure equilibrium between the magnetic pressure and the external pressure along the boundary of the tower, as $B_{z}^{2}/8\pi=p_{\rm ext}$, where $B_{z}$ is the vertical component of the field lines threading the BH (the red lines in Figure \ref{Fig_field_config}), which can be estimated by $B_{z}=\Phi_{\rm BH}/\pi R_{\rm m}^{2}$, with $\Phi_{\rm BH}$ being the magnetic flux threading the BH.
 The external pressure in a power-law form with the height as $p_{\rm ext}\propto Z^{-\eta}$ has been extensively discussed in previous works, with different values of $\eta$ explored by theoretical analysis and numerical calculations when studying jets (\citealt{Tchekhovskoy2008,Zakamska2008,Komissarov2009,Lyubarsky2009}). It has been demonstrated that the magnetic tower has a valid solution when $0\leq\eta\leq4$ if the ram pressure is not considered, otherwise the tower will not be collimated (\citealt{Lynden-Bell2003,Lynden-Bell2006,Sherwin2007}). 
By combining the relations $R_{\rm m}\propto p_{\rm ext}^{-1/4}$ and $p_{\rm ext}\propto Z^{-\eta}$, we obtain $R_{\rm m}\propto Z^{\eta/4}$, which implies that the geometry corresponds to a cylindrical shape when $\eta=0$, a conical shape when $\eta=4$; and other intermediate values of $\eta$ correspond to parabolic geometries. 

However, directly obtaining the external pressure distribution from observations is challenging. The geometry, on the other hand, can be observed through interferometry, such as in the M87 jet, which is suggested as $Z\propto R_{\rm m}^{1.73}$ (\citealt{Asada2012,Mertens2016}). Thus, we directly presume different geometries of the tower to mimic the effect of varying external pressures. The geometry is assumed to be expressed as
\begin{align}
Z/R_{\rm g}=\alpha(R_{\rm m}/R_{\rm g}-R_{\rm in}/R_{\rm g})^{\beta}, \label{shape}
\end{align}
where $\alpha$ and $\beta$ are parameters that control the geometry. In this expression, we simply consider the reconnection layer anchored at the disk's inner boundary, although other positions within the inner disk are also possible. Our calculations are not sensitive to this configuration. We set $R_{\rm in}=1R_{\rm g}$ for all our subsequent analyses, corresponding to the case of $a=1$.
When $Z$ is large, $R_{\rm m}\gg R_{\rm in}$ is satisfied and the tower's geometry follows a power-law form.
As an example, Figure \ref{Fig_Z_R}(a) plots the geometry of the tower boundary, with typical values of $\alpha$ and $\beta$ listed in Table~\ref{tab_parameter}. Here, we point out that larger values of $\alpha$ and $\beta$ correspond to a smaller radial extension in the geometry. It can be seen that the models from M1 to M9 demonstrate a transition from geometries with a mild radial extension to geometries with strong collimation. Also, the geometry is more sensitive to the power-law index $\beta$. 

With the shape of the tower's boundary, we can investigate the properties of the reconnection layer along this boundary.  
The reconnection layer is characterized by a curve length of $S_{\rm c}$ and a thickness of $W_{\rm c}$, as shown in Figure \ref{Fig_layer_config}. 
Furthermore, in the presence of plasmoids, the thickness of the reconnection layer would be determined by the size of the largest plasmoids, which has been observed in simulations and can be approximated as \citep{Sironi2016} 
\begin{align}
W_{\rm c}\sim0.2S_{\rm c}, \label{H_0.1S}
\end{align}

The size of the reconnection layer has been found to be at least $S_{\rm c}\sim 10R_{\rm g}$ and $W_{\rm c}\sim R_{\rm g}$ in simulations (\citealt{Ripperda2020}).

The relation between the height $Z_{\rm c}$ that the reconnection layer can reach and the curve length $S_{\rm c}$ is given by
\begin{align}
\int_{0}^{Z_{\rm c}}\sqrt{1+\left(\frac{dZ}{dR_{m}}\right)^{-2}}dZ=S_{\rm c}. \label{z_s}
\end{align}
Combining the shape of the tower's boundary given by Equations (\ref{shape}) and (\ref{z_s}), the reconnection layer would have a parabolic geometry with limited vertical and radial extensions.
Figure \ref{Fig_Z_R}b shows the relations between $S_{\rm c}$ and $Z_{\rm c}$ for $\alpha$ and $\beta$ listed in Table~\ref{tab_parameter}. The geometry with a relatively greater radial extension like M1 deviates from the linear dependence at low $Z_{c}$, while the other cases approximately follow a linear relation across all $Z_{\rm c}$.

\begin{figure*}[th!]
\centering
\includegraphics[height=6.5cm]{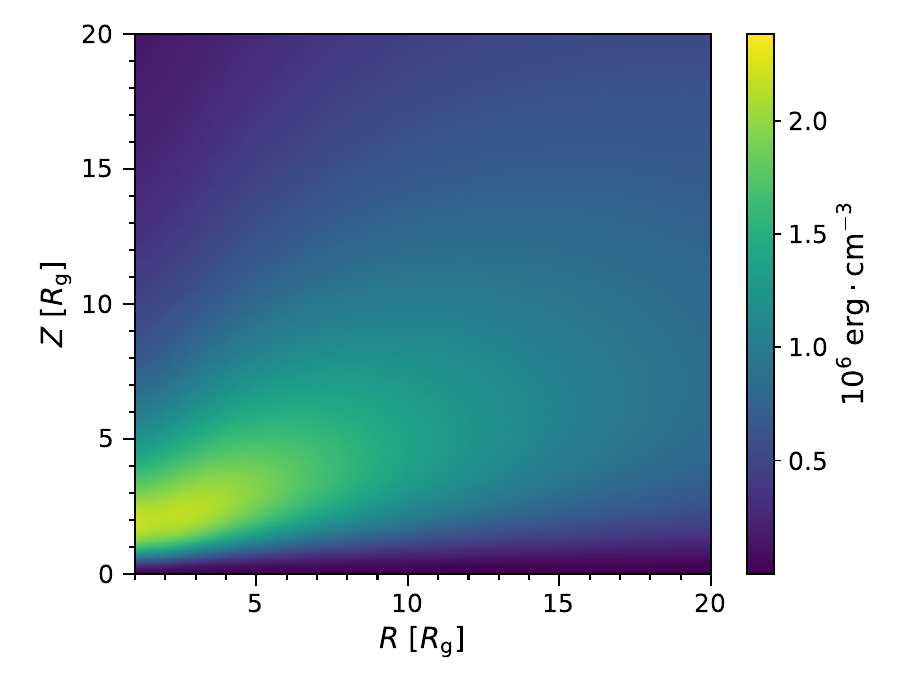}
\includegraphics[height=6.1cm, trim=0pt -10pt 0pt 15pt]{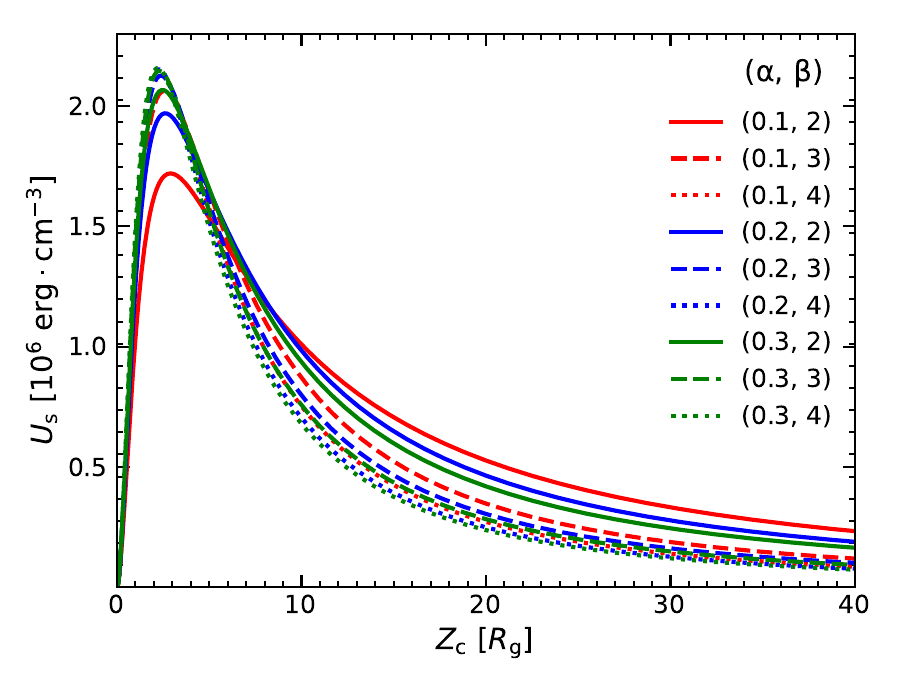}
\caption{Left: the energy density distribution of the soft photon field generated by the standard thin disk in the $R-Z$ plane for a BH-disk system with $m=10^{8}$, $ \dot{m}=1$, $R_{\rm in}=1R_{\rm g}$, and $R_{\rm out}=10^{4}R_{\rm g}$. For other $m$ and $\dot{m}$, the corresponding energy density of the photon field scales with a factor of $10^{8}/m\cdot\dot{m}$.
Right: the energy density distribution of the soft photon field varies with height for the different pairs of $\alpha$ and $\beta$ listed in Table~\ref{tab_parameter}.}
\label{Fig_Us_model}
\end{figure*}

\subsection{The Energy Density of the Soft Photon Field} \label{soft_photons}
It has been suggested that the soft photons for IC radiation in the hot corona originate from the cold accretion disk or the synchrotron radiation of relativistic electrons (\citealt{Beloborodov1999,Malzac2001,Beloborodov2017}). To calculate the IC radiation of the vertically extended reconnection layer, one needs the spatial distribution of the soft photons. Here, we focus on the soft photons emitted by the accretion disk, while the synchrotron photons are subject to self-absorption \citep{Beloborodov2017} and might be not important.

For a standard thin disk \citep{Shakura1973} with a central BH mass $M$, accretion rate $\dot{M}$, inner boundary $R_{\rm in}$ and outer boundary $R_{\rm out}$, the radiative energy flux at the disk surface is given by
\begin{align}
F(R)=\frac{3GM\dot{M}}{8\pi R^{3}}\left(1-\sqrt{\frac{R_{\rm in}}{R}}\right). \label{SSD_F}
\end{align}
If we assume isotropic radiation from the surface, the intensity is given by $I=F/2\pi$, which contributes to the energy density by $I/c$. The energy density of a static soft photon field at a spatial point $(R_{0},Z_{0})$ can be calculated by integrating over all solid angles, as
\begin{align}
U_{\rm s}(R_{0},Z_{0})=\frac{1}{c}\int_{0}^{2\pi}d\phi\int_{\theta_{\rm min}}^{\theta_{\rm max}}d\theta
\frac{F(R')}{2\pi}\cos\psi \sin\psi, \label{Us}
\end{align}
where 
\begin{align}
\theta_{\rm min}&=\arctan{\frac{R_{\rm in}}{Z_{0}}},\quad\theta_{\rm max}=\arctan\frac{R_{\rm out}}{Z_{0}}, \notag 
\end{align}
and
\begin{align}
&\cos\psi=\frac{Z_{0}}{\sqrt{Z_{0}^{2}+L^{2}}},\quad \sin\psi=\frac{L}{\sqrt{Z_{0}^{2}+L^{2}}}, \notag \\
&L=\sqrt{(R'\cos\phi+R_{0})^{2}+(R'\sin\phi)^{2}},\quad R'=Z_{0}\tan\theta. \notag 
\end{align}  
As an example, the left panel of Figure \ref{Fig_Us_model} shows the distribution of the energy density of the soft photon field in the $R-Z$ plane for the case of $m=10^{8}$, $\dot{m}=1, R_{\rm in}=1R_{\rm g}, R_{\rm out}=10^{4}R_{\rm g}$. In the radial direction, the energy density of the soft photon field is mainly concentrated at small radius, because the radiation flux decreases with radius, as given by Equation (\ref{SSD_F}). In the vertical direction, the energy density shows a trend of initially increasing and then decreasing, as plotted in the right panel of Figure \ref{Fig_Us_model}. This is because when it is sufficiently close to the disk plane, the radiation contribution from other radii tends toward zero, and when far enough from the disk plane, due to the attenuation of the solid angle, the radiation contribution from all radial positions rapidly decreases. Furthermore, the model with the most collimated geometry exhibits the most rapid decrease in energy density with height.
For a given dimensionless coordinate $(R_{0}/R_{\rm g},Z_{0}/R_{\rm g})$, Equation (\ref{SSD_F}) implies that the value of $U_{\rm s}(R_{0},Z_{0})$ is proportional to $\dot{m}$ but inversely proportional to $m$. 
Here, we note that the inner region of an accretion disk might transition to advection-dominated accretion flows (e.g., \citealt{Narayan1998}), therefore the radiation flux will deviate from Equation (\ref{SSD_F}). Nonetheless, we emphasize that the geometry and radiation properties of our model for the hot corona mainly depend on one characteristics of the radiative field: its energy density strongly decreases along the $z$-direction after reaching a certain height. This property is just due to the geometric effects of solid angles and is valid regardless of the states of accretion flows. Therefore, our analysis of the geometry and radiation of the hot corona is still qualitatively applicable for sources that cannot be characterized by a standard thin disk.

\subsection{The Dynamics of Plamoids in the Reconnection Layer} \label{dynamical_picture}
\begin{figure}
\centering
\scalebox{0.65}[0.65]{\rotatebox{0}{\includegraphics{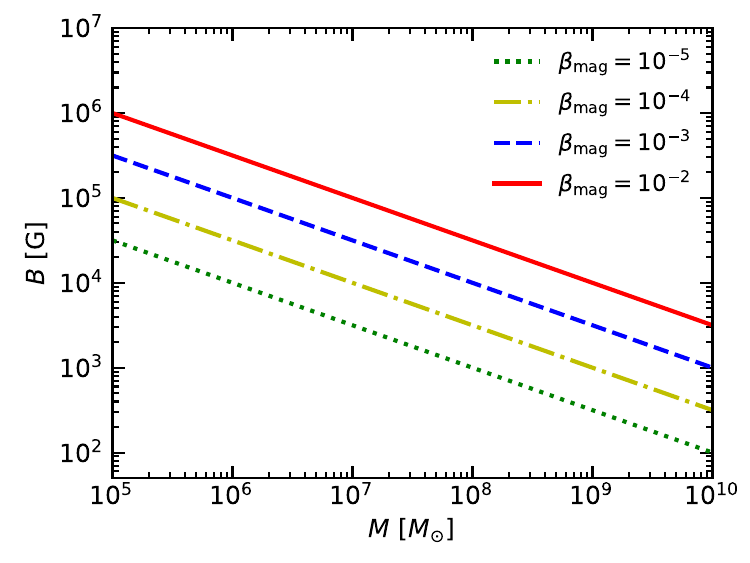}}}
\caption{The relation between the magnetic field strength at $R_{\rm in}$ and the central BH mass. The parameters $\alpha_{\rm vis}=0.1$ and $r=1$ are set. The solid, dashed, dash-dotted lines, and dotted lines correspond to the cases of $\beta_{\rm mag}=10^{-2}$, $10^{-3}$, $10^{-4}$, and $10^{-5}$, respectively.}
\label{Fig_B_m}
\end{figure}

An important parameter of the reconnection layer is the magnetization, defined as 
\begin{align}
\sigma_{\rm mag}=\frac{2U_{\rm B}}{\rho c^{2}}, \label{magnetization}
\end{align}
where $U_{\rm B}=B^{2}/8\pi$ is the energy density of the magnetic field and $\rho$ is the plasma mass density, primarily determined by the electron-positron mass, which is produced by the balance between pair creation of MeV photons and annihilation (\citealt{Fabian2015,Beloborodov2017}). The number density of $e^{\pm}$ can be calculated using the optical depth as
\begin{align}
n_{\pm}=\frac{\tau_{\rm T}}{\sigma_{\rm T}W_{\rm c}}, \label{tau}
\end{align}
where $\sigma_{\rm T}$ is the Thomson cross-section and $\tau_{\rm T}\sim1$ is the optical depth of the hot corona inferred by observations of AGNs \citep{Fabian2015,Wilkins2015}. As long as $\tau_{\rm T}$ does not deviate significantly from 1, the radiation power of the corona will not have a substantial difference in magnitude, as the number density of $e^{\pm}$ linearly depends on optical depth by Equation (\ref{tau}). Furthermore, considering that previous studies have shown that the optical depth does not change during X-ray variations \citep{Pal2023,Serafinelli2024}, we set $\tau_{\rm T}=1$ in all our subsequent analysis.
The plasma mass density can then be estimated as
\begin{align}
\rho\sim m_{\rm e}n_{\pm}. \label{rho}
\end{align}
The energy density of the magnetic field can be determined by assuming that the magnetic pressure is a fraction of the disk pressure
\begin{align}
  U_{\rm B}=\beta_{\rm mag}p_{\rm disk} \label{UB}.
 \end{align}
 In the inner region of a standard thin disk, the disk pressure is dominated by radiation pressure, given by 
 \begin{align}
 p_{\rm disk}\sim p_{\rm rad}=\frac{4\sigma T^{4}}{3c},
  \end{align}
 where $\sigma$ is the Stefan-Boltzmann constant. The disk temperature in a standard thin disk \citep{{Shakura1973}} is expressed as 
  \begin{align}
 T=6.3\times10^{7}(\alpha_{\rm vis} m)^{1/4}r^{-3/8}~\rm{K}\label{SSD_T},
   \end{align}
   where $\alpha_{\rm vis}$ is the viscosity parameter. 
 By combining Equations (\ref{magnetization}) - (\ref{SSD_T}), if $\alpha_{\rm vis}$, $\tau_{\rm T}$, $W_{\rm c}$ and $r$ are specified, one can calculate the corresponding magnetization for a given $\beta_{\rm mag}$. 
 The magnetization is not dependent on BH mass for a given $\beta_{\rm mag}$, because
 both the magnetic energy density and mass density are inversely proportional to the BH mass.
For a magnetically dominated corona, $\sigma>1$ should be satisfied. Considering $\alpha_{\rm vis}\sim 0.1$, $\tau_{\rm T}\sim 1$, $W_{\rm c}\sim 1R_{\rm g}$ and $r\sim 1$, this imposes a lower limit for $\beta_{\rm mag}$ of approximately $10^{-5}$. What's more, considering that the calculated radiation power of the reconnection layer depends on $\beta_{\rm mag}$, which we will quantitatively calculate below, the observed X-ray luminosity of AGNs constrains an upper limit for $\beta_{\rm mag}$ of approximately $10^{-2}$. Therefore, only a small fraction of the disk pressure is contributed by the magnetic pressure. We calculate and depict the magnetic field strength for various BH masses of $m=10^{5}-10^{10}$ and $\beta_{\rm mag}=10^{-5}-10^{-2}$ in Figure \ref{Fig_B_m}, with parameters $\alpha_{\rm vis}=0.1$, $r=1$. The magnetization corresponds to $\sigma_{\rm mag}\sim1-10^{3}$ for $\tau_{\rm T}=1$, $W_{\rm c}=1R_{\rm g}$.

 In the reconnection layer, a chain of plasmoids will form by tearing-unstable magnetic flux and particles will be accelerated by the strong tension of reconnected field lines. This process has been proposed and explored by numerical simulations \citep{Loureiro2007,Uzdensky2010,Sironi2016,Sironi2020,Sridhar2021,Sridhar2023}. It is important to study the dynamics of these plasmoids as it directly determines the radiation properties of the reconnection layer. We follow the theoretical analysis given in \cite{Beloborodov2017}.
  
 The plasmoids in the reconnection layer have diverse sizes, with the typical size of the smallest and largest plasmoids being
\begin{align}
W_{\rm min}&\sim R_{\rm L}~~{\rm and}~~W_{\rm max}\sim  W_{\rm c}, \label{omega_size}
\end{align}
respectively, where $R_{\rm L}\sim \sigma_{\rm mag} c/\omega_{\rm B}$ is the Larmor radius, $\omega_{\rm B}=eB/m_{\rm e}c$ is the electron gyration frequency, and $e$ represents the elementary charge. The size distribution of plasmoids follows a power law in a self-similar chain, as
\begin{align}
f(W)\propto W^{-1}, \label{DF}
\end{align}
where the normalization factor is given by $1/\ln(W_{\rm max}/W_{\rm min})$ and the power index -1 is recommended by \cite{Huang2012}.

In the reconnection layer, the plasmoids are subject to a push force exerted by the reconnected field lines. An approximate form of the push force per unit volume on a plasmoid with a size of $w$ is
\begin{align}
f_{\rm push}=\xi \frac{U_{\rm B}}{W}, \label{fpush}
\end{align}
where $\xi\sim0.1$ is taken in \citet{Beloborodov2017}. This expression is based on the assumption that the background magnetic field is uniform throughout the reconnection layer.

If one considers that the plasmoids move in a soft photon field, the IC scattering of relativistic electrons exerts a drag force on the plasmoids.
The drag force per unit volume acting on the plasmoids due to IC scattering is given by
\begin{align}
f_{\rm drag}=\beta\gamma^{2}U_{\rm s}\sigma_{\rm T}n_{\pm}, \label{fdrag}
\end{align}
where $\gamma=1/\sqrt{1-\beta^{2}}$, $\beta=v/c$, $v$ is the speed of the plasmoids, and $U_{\rm s}$ is the energy density of the soft-photon field, which we have quantitatively calculated in Section \ref{soft_photons}. 

For a plasmoid with a size of $W$, the push force remains approximately constant in the reconnection layer, whereas the drag force is positively correlated with its velocity. By combining Equation (\ref{tau}), (\ref{fpush}) and (\ref{fdrag}), the condition $f_{\rm push}=f_{\rm drag}$ will result in a maximum velocity $v_{\rm max}$, which satisfies
\begin{align}
\beta_{\rm max}\gamma_{\rm max}^{2}= \frac{\tau_{\star}}{\tau_{\rm pl}}~(\gamma_{\rm max}<\sqrt{\sigma_{\rm mag}}), \label{vmax}
\end{align}
where $\tau_{\star}\equiv \xi U_{\rm B}/U_{\rm s}$ and $\tau_{\rm pl}=\tau_{\rm T}W/W_{\rm c}$.
However, the maximum velocity of the plasmoids should be limited by the speed of the fast magnetosonic waves (\citealt{Lyubarsky2005,Sironi2016}), which corresponds to the Lorentz factor $\gamma\approx \sqrt{\sigma_{\rm mag}}$. 
As an example, if $\tau_{\star}=1$ and $\sigma_{\rm mag}=10$, for plasmoids with $W<0.1W_{\rm c}$, the Lorentz factor is limited to $\gamma_{\rm max}=\sqrt{\sigma_{\rm mag}}\approx3.3$.
Thus, large plasmoids will be constrained by Equation (\ref{vmax}) to reach only mild relativistic velocities, while small plasmoids will be accelerated to relativistic velocities but limited by the Lorentz factor $\sqrt{\sigma_{\rm mag}}$.
What is more, given that the energy density of the soft-photon field varies with height (see Figure \ref{Fig_Us_model}), the maximum velocity of a plasmoid with a given size is different at various heights.
With a large Lorentz factor of $\gamma_{\rm max}\sim 30$ for the $\sigma_{\rm mag}=10^{3}$ case, in a standard thin disk with a characteristic high temperature of $T\sim10^{6}\rm\,K$, the peak wavelength of the blackbody radiation is $\lambda_{\rm peak}\sim 30\,\text{\AA}$, which still satisfies the condition $\gamma_{\rm max}hc/\lambda_{\rm peak}\ll m_{\rm e}c^{2}$, where $h$ is the Planck constant and $m_{\rm e}$ is the mass of the electrons. Therefore, the radiation field calculated in Section \ref{soft_photons} can be safely considered as a soft-photon field for IC scattering, as almost all photons are in the Thomson regime. 

The gravitational force per unit volume acting on the largest plasmoids can be estimated by
\begin{align}
 f_{\rm grav}\sim \frac{GMm_{\rm e}n_{\pm}}{R_{\rm in}^{2}}. \label{f_grav}
\end{align}
Combining Equations (\ref{fpush}) - (\ref{f_grav}), one can find that for a plasmoid with velocity $0\leq v \leq v_{\rm max}$,
\begin{align}
f_{\rm push}\geq f_{\rm drag}\gg f_{\rm grav} \label{force_relation},
\end{align}
is always satisfied. Thus, one can ignore the gravitational force when considering the dynamics of the plasmoids. When $v=v_{\rm max}$, the equal sign in the relation of Equation (\ref{force_relation}) holds true. 

For a plasmoid with zero initial velocity, the characteristic distance required for it to be accelerated to $v_{\rm max}$ is determined by
\begin{align}
\Delta S_{1}=n_{\pm}m_{\rm e}\int_{0}^{v_{\rm max}}\frac{v}{f_{\rm push}-f_{\rm drag}(v)}dv.
\end{align}
If the plasmoid moves outside (vertically) the reconnection layer, the push force should vanish in an unreconnected force-free field. The characteristic distance required for it to be decelerated to $v=0$ is calculated by
\begin{align}
\Delta S_{2}=n_{\pm}m_{\rm e}\int_{0}^{v_{\rm max}}\frac{v}{f_{\rm drag}(v)}dv.
\end{align}
Our calculations find that $\Delta S_{1}\ll S_{\rm c}$ and $\Delta S_{2}\ll S_{\rm c}$.
 
Based on the above analysis, within the reconnection layer, the plasmoids will rapidly reach their maximum velocity. Once moving outside (vertically) the reconnection layer, the plasmoids will quickly be decelerated and no longer contribute to the radiation, thus the primary radiation region should be within the reconnection layer.
Therefore, for simplicity in the calculations, we propose a simple yet reasonable dynamical picture for the plasmoids chain---namely, they are distributed in size according to Equation~(\ref{DF})
and move at their maximum velocities of $v_{\rm max}$ along the reconnection layer. 

\subsection{The Radiation of Plasmoids Chain} \label{radiation}
With the geometry of the reconnection layer in Section \ref{geometry}, the energy density of the soft photon field in Section \ref{soft_photons}, and the dynamical picture of the plasmoids in Section \ref{dynamical_picture}, we are now able to estimate the power of the IC scattering of electrons in the plasmoid chain, which we expect to be the primary source of the X-rays produced by the hot corona.

The IC-scattering power of a single electron \citep{Rybicki1979} in plasmoids with size $W$ and velocity $v_{\rm max}$ is given by
\begin{align}
P_{\rm IC}(W,Z)=\frac{4}{3}\sigma_{\rm T}c\gamma_{\rm max}^{2}(W,Z)\beta_{\rm max}^{2}(W,Z)U_{\rm s}(Z), \label{P_single}
\end{align}
where $\beta_{\rm max}(W,Z)$ and $\gamma_{\rm max}(W,Z)$ are determined by Equation (\ref{vmax}) and limited by the
speed of the fast magnetosonic waves, while $U_{\rm s}(Z)$ is calculated by Equation (\ref{Us}).
The probability distribution of a radiating electron originating from a plasmoid with size $W$ is described by Equation (\ref{DF}), thus the contribution of IC radiation from all the electrons in the annular cross section of the reconnection layer at height $Z$ can be estimated as
\begin{align}
P_{\rm cir}(Z)\sim 2\pi R_{\rm m}(Z)n_{\pm}\int W f(W) P_{\rm IC}(W,Z)dW, \label{P_cir}
\end{align}
where $R_{\rm m}(Z)$ reflects the geometry of the reconnection layer as Equation~(\ref{shape}). In this expression, we assume that the plasmoids fill up the reconnection layer at each height, while in reality, a filling factor is expected, which leads to a correction factor on the order of unity in Equation (\ref{P_cir}). We ignore this factor for simplicity. 
Once $S_{\rm c}$, $\tau_{\rm T}$ are given, $W_{\rm c}$ is estimated by Equation~(\ref{H_0.1S}), and $n_{\pm}$ is derived by Equation (\ref{tau}). Here, considering that $W_{\rm c}\sim R_{\rm g}$, we neglect the small variations in energy density of the soft photon field along the radial direction (see Figure ~\ref{Fig_Us_model}).
We then integrate all contributions from the entire reconnection layer to obtain the total radiation power as
\begin{align}
P_{\rm tot}=\int_{0}^{S_{\rm c}}P_{\rm cir}(Z_{\rm c})dS, \label{P_tot}
\end{align}
where $S$ is the parabola described by Equation (\ref{shape}) and $Z_{c}(S)$ is given by Equation (\ref{z_s}).
By combining with Equation (\ref{z_s}), we define the characteristic height of the AGN hot corona as
\begin{align}
H_{\rm cor}=\frac{\int_{0}^{S_{\rm c}}P_{\rm cir}(Z_{\rm c})Z_{\rm c}dS}{\int_{0}^{S_{\rm c}}P_{\rm cir}(Z_{\rm c})dS}.  \label{H_cor}
\end{align}
This definition represents the effective height of the vertically extended radiative corona, which we associate with the height of the point-like hot corona in the lamp-post model.

\begin{figure}
\centering
\scalebox{0.6}[0.6]{\rotatebox{0}{\includegraphics{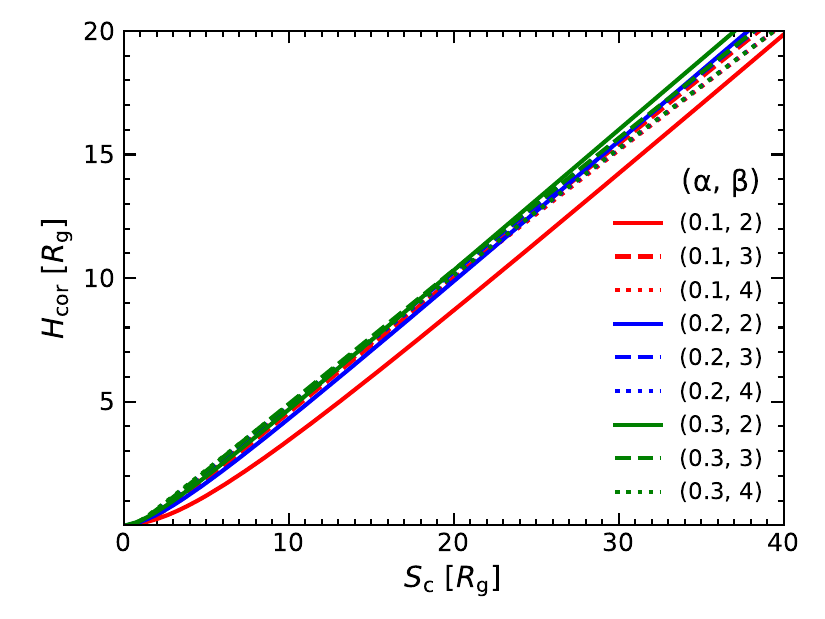}}}
\caption{The relation between the size of the reconnectin layer and the defined height of the hot corona for all pairs of $\alpha$ and $\beta$ listed in Table~\ref{tab_parameter}. The BH-disk system has $m=10^{8}$, $\dot{m}=1$, $\tau_{\rm T}=1$, and $\beta_{\rm mag}=10^{-3}$.}
\label{Fig_Sc_Hcor}
\end{figure}

\begin{figure*}[t]
\centering
\includegraphics[height=6.5cm]{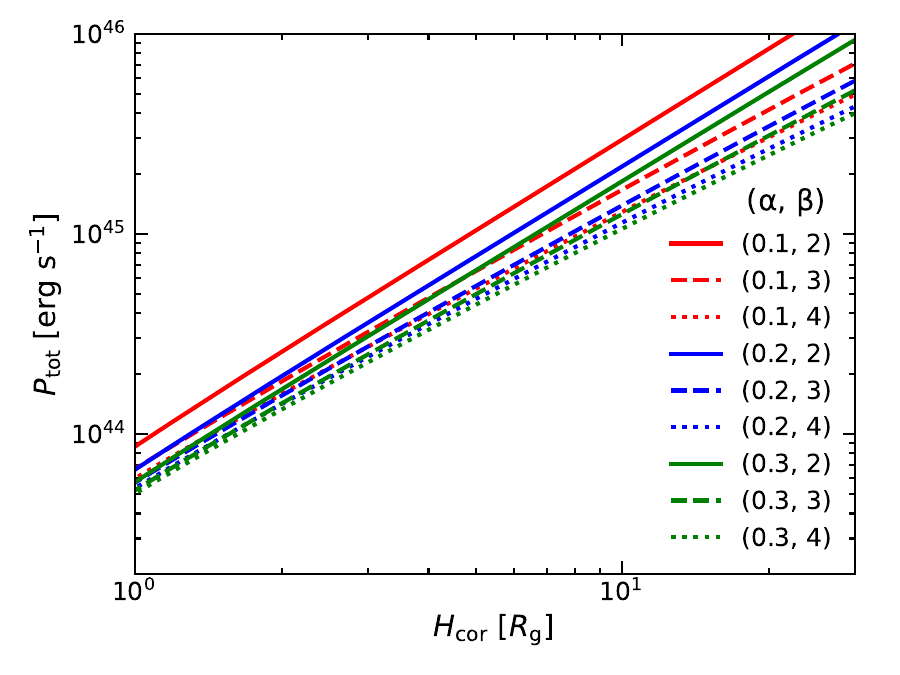}
\includegraphics[height=6.5cm]{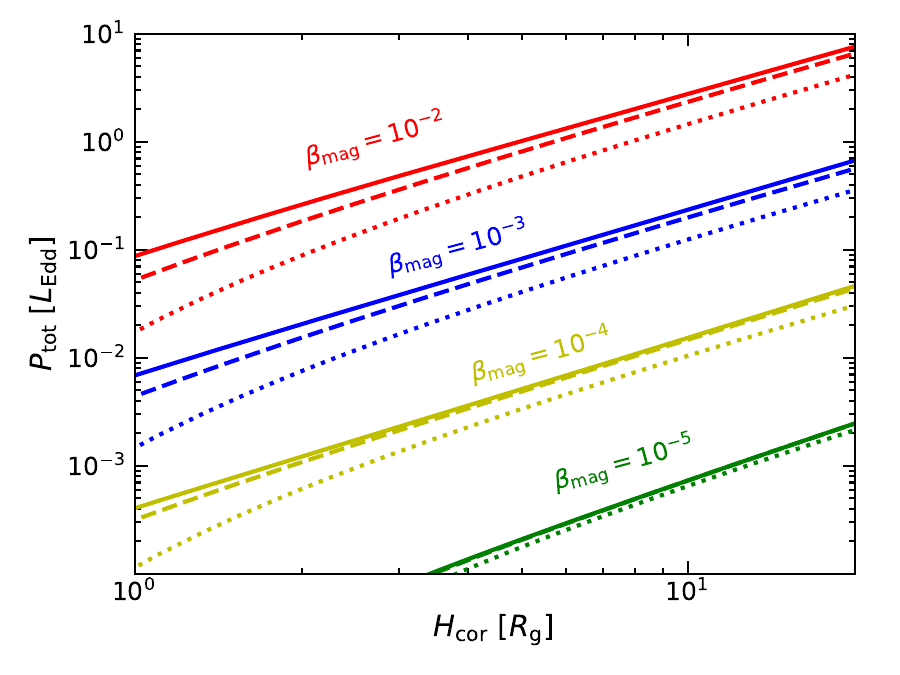}
\caption{Left: the relation between the IC radiation power and the defined height of the hot corona for all pairs of $\alpha$ and $\beta$ listed in Table~\ref{tab_parameter}. The BH-disk system has $m=10^{8}$, $\dot{m}=1$, $\tau_{\rm T}=1$, and $\beta_{\rm mag}=10^{-3}$. Right: the relation between the IC radiation power and the defined hot corona's height for $(\alpha=0.1, \beta=2)$ for different $\beta_{\rm mag}$ and $\dot{m}$. The total power is in units of the Eddington luminosity. The solid, dashed, and dotted lines correspond to $\dot{m}=1$, 0.1, and 0.01, respectively. The green, yellow, blue, and red colors represent $\beta_{\rm mag}=10^{-5}, 10^{-4}, 10^{-3}, and 10^{-2}$, respectively.}
\label{Fig_Ptot_Hcor}
\end{figure*}

In Figure \ref{Fig_Z_R}, we show the geometry of the reconnection layer with different combinations of the parameters $\alpha$ and $\beta$ in Equation (\ref{shape}). We test a wide range of the two parameters and discard those combinations with excessive radial extension, as they do not align with observations. Specifically, the hot corona should locate within a few tens of gravitational radii from the central BH. Some typical sets of $\alpha$ and $\beta$ are shown in Table \ref{tab_parameter}. 
By combining Equations (\ref{shape}) - (\ref{vmax}) and (\ref{P_single}) - (\ref{H_cor}), we can calculate the corresponding IC-scattering power and the defined corona height once the parameters $\beta_{\rm mag}$, $S_{\rm c}$, and $\tau_{\rm T}$ are specified.
 As an example, by fixing the parameter $\beta_{\rm mag}=10^{-3}$ and adjusting the size of the reconnection layer, Figure \ref{Fig_Sc_Hcor} illustrates that the defined corona height exhibits an approximately linear relation with the size of the reconnection layer, and the left panel of Figure \ref{Fig_Ptot_Hcor} shows the relation between the IC-scattering power and the defined corona height for $m=10^{8}$, $\dot{m}=1$, and $\tau_{\rm T}=1$. 
The IC-scattering power roughly increases linearly with the corona height, as there are more electrons and positrons in a larger reconnection layer via Equations (\ref{H_0.1S}) and (\ref{tau}). 
Furthermore, the left panel of Figure \ref{Fig_Ptot_Hcor} demonstrates that reconnection layers with different geometries do not show significant differences in IC-scattering-power magnitude at the same height. This is due to the fact that in the inner region of the soft-photon field, the energy density provided by Equation (\ref{Us}) does not decline significantly along the $r$-direction, as shown in Figure \ref{Fig_Us_model}. Nevertheless, from Figure \ref{Fig_Z_R} and the left panel of Figure \ref{Fig_Ptot_Hcor}, one can find that the reconnection layer that extends farther along the $r$-direction has a lower height with the same radiation power. 
This is because the energy density of soft photon field in a reconnection layer with strong collimated geometry decreases more rapidly with height, as illustrated in Figure \ref{Fig_Us_model}, requiring a larger vertical extension to reach the same radiation power compared to a reconnection layer with a larger radial extension.

Since all geometries of the reconnection layer produce similar magnitudes of the total power, as in the left panel of Figure \ref{Fig_Ptot_Hcor}, we utilize the model M1 ($\alpha=0.1$, $\beta=2$) with different values of $m$, $\dot m$ and $\beta_{\rm mag}$ to demonstrate the dependence of the total power on the BH mass, accretion rate, and magnetic field strength. 
For a given $\beta_{\rm mag}$, since both $U_{\rm B}$ and $U_{\rm s}$ are inversely proportional to the BH mass via Equations (\ref{Us}) and (\ref{magnetization}) - (\ref{rho}), the velocity distribution of the plasmoids remains unchanged with varying BH mass. One can conclude that the total radiation power calculated in Section \ref{radiation} can be scaled by the BH mass, although $W_{\rm min}$ may result in a small deviation, as it is not linearly dependent on the BH mass. Hence, we use the Eddington luminosity as the unit to express the total radiation power. 
For different $\dot m$ and $\beta_{\rm mag}$ values, we depict the relation of the total radiation power with corona height in the right panel of Figure \ref{Fig_Ptot_Hcor}. It is observed that a smaller $\dot m$ corresponds to a lower radiation power, which is due to the relation that $U_{\rm s}\propto \dot m$ via Equation (\ref{Us}). But since the maximum velocities of large plasmoids are enhanced by lower $U_{\rm s}$ in Equation (\ref{vmax}), the radiation does not exhibit linear dependence on the accretion rate. 
Moreover, the radiation power exhibits a strong dependence on $\beta_{\rm mag}$, which can be understood as a stronger magnetic field leading to more magnetic energy being transferred to particle radiation. The total power typically ranges from $10^{42}$ to $10^{46}~\rm{erg~s^{-1}}$ for $m\sim10^{6}-10^{9}$, $\dot{m} \sim 0.01-1$, and $\beta_{\rm mag}\sim10^{-4}-10^{-3}$, which represent the typical conditions for AGNs, approximately consistent with the observed X-ray luminosity of AGNs \citep{Mateos2015}. The corresponding magnetic field strength at the innermost disk ranges from $10^{3}-10^{5}~\rm{G}$, as shown in Figure \ref{Fig_B_m}, which also agrees with the spectropolarimetry observations  \citep{Piotrovich2021}.

\begin{figure}[t!]  
\centering 
\scalebox{0.6}[0.6]{\rotatebox{0}{\includegraphics{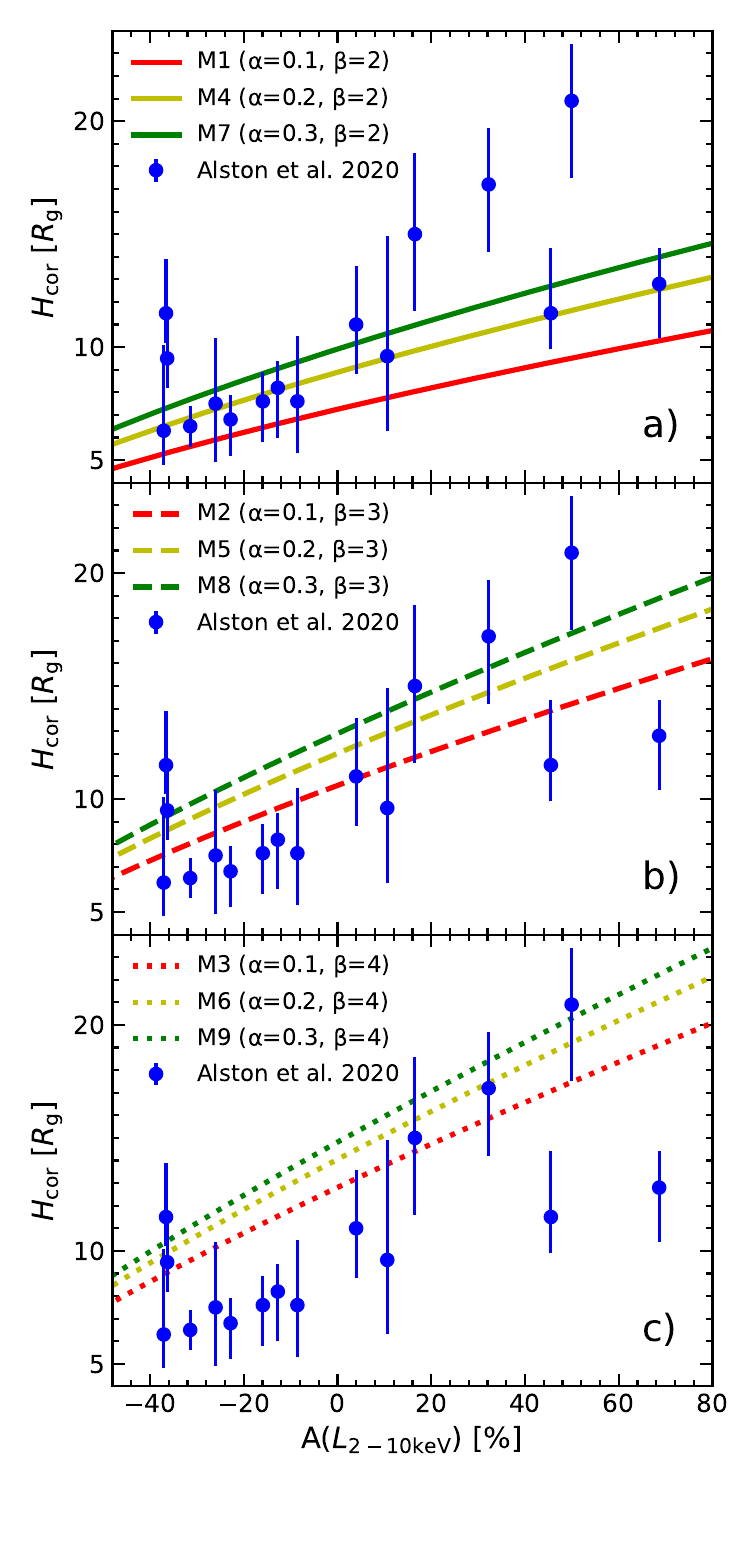}}}
\caption{The relation between the height of the hot corona and the relative amplitude in the 2-10 keV band for IRAS 13224-3809. The blue points represent the observations by \citealt{Alston2020}. The lines represent our theoretical calculations using the sets of $\alpha$ and $\beta$ listed in Table~\ref{tab_parameter}. From top to bottom are the results of (a) M1, M4, M7  ($\beta=2$), (b) M2, M5, M8 ($\beta=3$), and (c) M3, M6, M9 ($\beta=4$), respectively.}
\label{Fig_IRAS_fit}
\end{figure}

\subsection{Application to IRAS 13224-3809} \label{application}
Reverberation mappings in X-rays offer an observational way of probing the height of the hot corona. 
In this section, we apply the above analyses to IRAS 13224-3809, with extensive X-ray reverberation mapping observations \citep{Alston2020}, this being a radio-quiet AGN with extreme spin parameter $a\sim 1$, BH mass $m\sim1.9\times 10^{6}$, and high accretion rate $\dot{m}\sim1$ (\citealt{Alston2019,Alston2020}). One can consider its X-ray emission completely from the hot corona, without the contribution of the jet, and its disk is more likely to be a standard thin disk due to the high accretion rate. We set $a=1$, $m=1.9\times10^{6}$, and $\dot{m}=1$ in our calculations. Below, we compare the relation between the corona height and 2-10 keV luminosity predicted by our theoretical model against the observations from X-ray reverberation mapping by \citet{Alston2020}.

To determine the radiation power within a specific energy band from the total radiation power, one needs the energy's lower and upper limits, as well as the spectral shape. The upper energy limit should be determined by the electron temperature in the hot corona with a range of $kT_{\rm e}\sim50-200\,\rm{keV}$ for various sources (\citealt{Beloborodov1999,Fabian2015,Wilkins2015}). This upper limit does not significantly affect the power in a specific energy band, as the power-law spectrum decays rapidly at high energies. The lower limit of the energy depends on the energy of the upscattering soft photons $\gamma_{\rm max}^{2}kT_{\rm s}$.
However, it is challenging to determine a specific value of $kT_{\rm s}$, as the soft-photon field is derived from integrating the radiation over the entire accretion disk in Section \ref{soft_photons}. In \citet{Beloborodov2017}, $kT_{\rm s}=10^{-3}m_{\rm e}c^{2}$ is assumed to obtain the spectrum from Monte Carlo simulations.
Considering the above complexity, we alternatively use the relative amplitude to describe the variation in luminosity, defined as,
\begin{align}
A(L)=\frac{L-\bar{L}}{\bar{L}}\times 100\%, 
\end{align}
where $L$ is the time-dependent luminosity and $\bar{L}$ is the time-average luminosity.

The hard-X-ray radiation flux of IRAS 13224-3809 follows a power-law distribution, with the photon index $\Gamma$ being time-dependent but not vary significantly \citep{Chiang2015}. It is reasonable to use a fixed shape to approximately describe the spectrum over the entire time period. With the constant lower and upper energy limits, one can derive that the 2-10 keV luminosity $L_{\rm 2-10keV}$ being proportional to the total power $P_{\rm tot}$, leading to the relation
\begin{align}
A(L_{\rm 2-10keV})=A(P_{\rm tot}). \label{amp}
\end{align}
As such, for IRAS 13224-3809, we can utilize the relative amplitude of the total power from our models to represent the variation in 2-10 keV luminosity from observations, without the need to convert the total power into 2-10 keV luminosity.

We first need to constrain the parameter $\beta_{\rm mag}$, since the radiation power in our model exhibits a strong dependence on it. Observations indicate that $\bar{L}_{\rm 2-10keV}=4.0\times10^{42}\rm{erg~s^{-1}}$, with a time-averaged corona height $\bar{H}_{\rm cor}=10.6R_{\rm g}$ and a time-averaged photon index $\bar{\Gamma}=2.4$ for IRAS 13224-3809 (see the Extended Data Figure 2 in \citealt{Alston2020}). If one simply considers that the X-ray energy ranges from 0.3-100~keV, the power-law spectrum yields a conversion factor of 0.25 when converting the total radiation power to the luminosity in 2-10 keV band, resulting in a rough estimation for the average total radiation power $P_{\rm tot}\sim1.6 \times 10^{43}~\rm{erg~s^{-1}}$.
We alternately select a reconnection layer with the geometry of set M2 in Table~ \ref{tab_parameter}, which represents the intermediate case in terms of geometry in Figure \ref{Fig_Z_R}a. By adjusting $\beta_{\rm mag}$, we ensure that the curve calculated in Section \ref{radiation} aligns with the point corresponding to the observed average luminosity and height of the hot corona, resulting in $\beta_{\rm mag}=5.1\times10^{-4}$.
With this $\beta_{\rm mag}$, and varying the size of the reconnection layer, we calculate the corresponding relative amplitude in the total power and corona height for all pairs of $\alpha$ and $\beta$ listed in Table~\ref{tab_parameter}. In Figure \ref{Fig_IRAS_fit}, we compare the relation between $A(L_{\rm 2-10keV})$ and $H_{\rm cor}$ calculated by our theoretical models against the observations by \citet{Alston2020}.

Figure \ref{Fig_IRAS_fit}(a) illustrates that the predicted relations of parameter sets M1, M4, and M7 generally align with the observations for $H_{\rm cor}<13R_{\rm g}$ in both amplitude and slope, while the data points with larger heights ({$\sim15-20R_{\rm g}$}) cannot be reproduced by these three models. In Figure \ref{Fig_IRAS_fit}(b), the parameter sets M2, M5, and M8 yield higher positions and stepper slopes in the height-luminosity relation, which match the data points with relatively larger heights ($\sim 15R_{\rm g}$) but fail to reproduce those data points with the lowest heights ($\sim 6R_{\rm g}$).
In Figure \ref{Fig_IRAS_fit}(c), the parameter sets M3, M6 and M9 yield the steepest slopes and can only reproduce the data points with the largest heights ($\sim 20R_{\rm g}$). These results overall demonstrate that the corona geometry with a small radial extension ($\beta\sim3-4)$ better reproduces the points at relatively larger heights, while the corona geometry with a larger radial extension ($\beta\sim2$) better reproduces the observation points at relatively lower heights. However, when the corona geometry's radial extension is excessively large (the red solid line in Figure \ref{Fig_IRAS_fit}), the resulting curve's slope is too small to fit the observations. In summary, the observation points can be generally fitted by our theoretical models, with a corona having a mild radial extension preferred for most observation points ({$\beta\sim3$}).
Moreover, it seems that a single corona shape cannot account for all data points, thus it is plausible that the hot corona may vary in geometry over the observation period for IRAS 13224-3809. During this period, the geometry of the corona exhibits a mild radial extension for most of time but a strongly collimated geometry occasionally. These geometric changes may be caused by variations in external pressure or the magnetic flux threading the BH, as discussed in Section \ref{geometry}. 

It is worth stressing that in the above analyses, we have chosen a reconnection layer with the geometry of M2 to match the average corona height and luminosity from observations, resulting in $\beta_{\rm mag}=5.1\times10^{-4}$, to ensure that the curve of M2 passes through the point (0,10.6) in Figure \ref{Fig_IRAS_fit}(b).
We can also adjust the value of $\beta_{\rm mag}$ to make the curves of other parameter sets listed in Table \ref{tab_parameter} fit the same point. This adjustment leads to minor differences from the adopted value for M2, such as $\beta_{\rm mag}=3.1\times10^{-4}$ for M1 and $\beta_{\rm mag}=7.6\times10^{-4}$ for M9. 
The variation in $\beta_{\rm mag}$ will result in minor changes to the horizontal positioning of the curve in Figure \ref{Fig_IRAS_fit} for each parameter set. However, this adjustment will not alter the slopes of the curves or the relative offsets between the curves. Therefore, our main results are not affected by the choice of different values of $\beta_{\rm mag}$. 

\section{Summary and Discussions} \label{discussion}
A new interpretation is presented for the hot corona in AGNs, which is linked to a chain of plasmoids in a reconnection layer with parabolic geometry. This scenario replicates the major properties of the hot corona in observations.
We first introduce a simplified magnetic field configuration that leads to the formation of the plasmoid chain within a reconnection layer along the magnetic tower's boundary.
Subsequently, we analyze the geometry of the reconnection layer, the energy density of the soft-photon field generated by the accretion disk, and the dynamical behavior of the plasmoids. Finally, we establish a relation between the radiation power of the reconnection layer and the corona height. 
The application of this scenario to IRAS 13224-3809 demonstrates a good fit to observations regarding the relation between the luminosity in X-ray and corona height, implying that the geometry of the hot corona has a mild radial extension and may evolve with time.
Moreover, the calculated X-ray luminosity in our model ranges from $10^{42}$ to $10^{46}~\rm{erg~s^{-1}}$ for $m\sim10^{6}-10^{9}$ and $\beta_{\rm mag}\sim10^{-4}-10^{-3}$, which is approximately consistent with the observed X-ray luminosity of AGNs \citep{Mateos2015}. The corresponding strength of the magnetic field ranges from $10^{3}$ to $10^{5}~\rm{G}$, which also agrees with the field strength estimated by observations from optical spectropolarimetry \citep{Piotrovich2021}.
As a result, the X-ray radiation from the hot corona of AGNs is attributed to the IC scattering of relativistic electrons in the plasmoids chain. 

The main parameters of our model are as follows. Two parameters $\alpha$ and $\beta$ in Equation (\ref{shape}) determine the shape of the reconnection layer, and the parameter $\beta_{\rm mag}$ determines the magnetic field strength. 
There are also several parameters such as the optical depth $\tau_{\rm T}$, the mass of the central BH $m$, and the accretion rate $\dot{m}$, with their typical values assigned according to observations. Once these parameters are specified, the relation between the X-ray radiation power and hot-corona height can be quantitatively established. The typical values of these parameters have been investigated and $\beta_{\rm mag}\sim5\times10^{-4}$ is chosen for IRAS 13224-3809, to align with the average luminosity and corona height from observations. 

Below, we conclude with remarks on two important but still open issues, which need further exploration in future works.

\subsection{Variations in the Size of the Reconnection Layer}

By adjusting the size of the reconnection layer, we have successfully fitted the observations of IRAS 13224-3809 with our theoretical models, by attributing the X-ray emission from the hot corona to the IC radiation of the reconnection layer. 
The height of the corona of IRAS 13224-3809 varies between $6.3R_{\rm g}$ and $20.9R_{\rm g}$ from its average value of $10.6R_{\rm g}$. In our models, the corona height $H_{\rm cor}$ shows an approximately linear dependence on the size of the reconnection layer $S_{\rm c}$ (Figure \ref{Fig_Sc_Hcor}), suggesting that the size of the reconnection layer needs to vary by approximately a factor of 2 to account for the observed heights in IRAS 13224-3809.
The variations in the size of the reconnection layer may result from the advection of magnetic loops, as discussed in Section \ref{field_rec}, which occurs on the viscous timescale
given by $ t_{\rm vis}\sim (R/H)^{2}/\alpha\Omega_{\rm k}$, where $H$ is the disk height and $\Omega_{\rm k}$ is the Keplerian angular velocity. In the innermost region of the disk, this timescale corresponds to days for IRAS 13224-3809, which is consistent with the observed timescale of variations in corona height and luminosity. 

From the perspective of energy conservation, one can imagine that the reconnection layer stays in an ideal equilibrium state, where the magnetic energy carried by accretion flow is efficiently converted into the radiation of particles. 
A sudden increase or decrease in the number of magnetic loops per unit time from the average value carried by the accretion flow will result in either an injection or deficit of the magnetic energy for the reconnection layer. It is reasonable to expect that the reconnection layer would adjust its characteristics, including its size, to achieve a new equilibrium state. This adaptation of the reconnection layer to the background conditions could occur on a timescale of MHD processes, which are typically much shorter than the viscous timescale of the accretion disk. According to the calculations in Section \ref{radiation}, it can be approximated that the output power from the reconnection layer is proportional to its size, expressed as $P_{\rm tot}\propto S_{\rm c}$. Therefore, it is expected that the reconnection layer will adjust its size in response to an injection or deficit of input power to maintain energy conservation.
In an extreme scenario, where no magnetic energy is provided by accretion, the reconnection layer would disappear, corresponding to the case of $S_{\rm c}=0$.  

However, we stress that the actual situation is likely more intricate than the simple qualitative analysis above. 
In our models, the magnetic energy density $U_{\rm B}$ is determined by $\beta_{\rm mag}$, which we have assumed to be constant, but in reality, it should fluctuate along with accretion with the magnetic loops. According to Section \ref{dynamical_picture}, the background magnetization and the dynamical behavior of plasmoids are expected to be influenced by the accretion process. The reconnection rate $\beta_{\rm rec}=v_{\rm rec}/c$, where $v_{\rm rec}$ is the reconnection speed, is an output parameter in reconnection process. It is commonly suggested that $\beta_{\rm rec}\sim 0.1$, based on simulations and observations (see, e.g., \citealt{Lyubarsky2005,Sironi2016,Cassak2017}), but it may also vary in response to changes of the background conditions. Simulations have indicated that the reconnection rate weakly depends on the background magnetization (\citealt{Sironi2016}). Therefore, the impact of the magnetization or other background conditions on the properties of the reconnection layer---such as its size, the width of the largest plasmoids, and the reconnection rate---remains unknown and requires further investigations through simulations.

\subsection{Verification of a Magnetic Reconnection Origin for Variations in X-Rays}
Below, we discuss a possible scenario for variations in X-rays by magnetic reconnection, following the analyses in \citet{Beloborodov2017}, and propose a method for verification. 
There are two components in the reconnection layer that contribute to radiation: the high-energy particles near the reconnection points and the plasmoids. The high-energy particles have a high Lorentz factor of hundreds, which emit both synchrotron and IC radiation, but only a small fraction of $\sim0.1$ of the magnetic energy is dissipated by them \citet{Beloborodov2017}, thus we do not consider their X-ray emission contribution in Section \ref{radiation}. In comparison, the particles in plasmoids have a Lorentz factor that is approximately 2 orders of magnitude lower than that of high-energy particles, with their synchrotron radiation being suppressed by synchrotron self-absorption, thus most of the magnetic energy is dissipated by IC emission in plasmoids. It can be concluded that the radio emission mainly originates from the high-energy particles, while the X-rays mainly comes from plasmoids. 
The power of synchrotron radiation is much lower than that of IC radiation. Nevertheless, the radio emission by synchrotron radiation of high-energy particles could be an important tool for verifying the magnetic reconnection origin of variations in X-rays.

The changes in the sizes of the reconnection layers are accompanied by corresponding variations in the number of reconnection points (\citealt{Uzdensky2010}). 
If the reconnection points have an increase, more particles will be accelerated near the reconnection points to become high-energy particles. Initially, the synchrotron radiation of these newly accelerated high-energy particles leads to the variations in radio emission, while their IC radiation will not cause a significant change in the X-ray emission, due to their minor contribution compared to plasmoids (\citealt{Beloborodov2017}).
Subsequently, the MeV photons from the IC radiation of high-energy particles will create secondary pairs, providing the injection of particles into young plasmoids, which merge and grow in size to form a plasmoid chain in a reconnection layer with a larger size (\citealt{Beloborodov2017}). During this process, the IC emission of these newly formed plasmoids will result in the observed variation in X-ray emission. Therefore, one may expect a correlated radio and X-ray variation, with the radio leading the X-ray, with the time lag arising from the process of formation from high-energy particles to radiative plasmoids.
In observations, the phenomenon of correlated radio and X-ray flares occurring, with radio preceding X-ray variations, has been observed in stellar corona (\citealt{Neupert1968}), and it has been suggested as a diagnostic method for the magnetic reconnection origin of the corona in radio-quiet AGNs \citep{Panessa2019}. By combining the analyses above, it can also serve as a verification for the scenario of our proposed X-ray variations, which requires the monitoring of radio and X-ray emissions in AGNs.
\section*{Acknowledgements}
We thank Luis C. Ho for useful discussions. We acknowledge financial support from the the National Natural Science Foundation of China (NSFC; 11991050 and 12333003) and the National Key R\&D Program of China (2021YFA1600404 and 2023YFA1607904). Y.R.L. acknowledges financial support from the NSFC through grant No. 12273041 and from the Youth Innovation Promotion Association CAS.

\bibliography{ref}
\bibliographystyle{aasjournal}
\end{document}